\documentclass[preprint,pra,aps,floatfix,showpacs]{revtex4-1}
\usepackage{epsfig,graphicx,tabularx,bm}

\begin{document}
\title{Bose-Hubbard model in a ring-shaped optical lattice with high filling factors }
\author{H. M. Cataldo and D. M. Jezek}
\affiliation{IFIBA-CONICET
\\ and \\
Departamento de F\'{\i}sica, FCEN-UBA
Pabell\'on 1, Ciudad Universitaria, 1428 Buenos Aires, Argentina}
\date{\today}
\begin{abstract}
The high-barrier quantum tunneling regime of a Bose-Einstein condensate confined in a ring-shaped optical 
lattice is investigated.
By means of a change of basis transformation, connecting
the set of `vortex' Bloch states and a Wannier-like set of localized wave functions,
we derive a generalized Bose-Hubbard Hamiltonian. 
In addition to the usual hopping rate terms, such a Hamiltonian takes into account
interaction-driven tunneling processes, which are shown to play a principal role at high filling
factors, when the standard hopping rate parameter turns out to be 
negative. 
By calculating the energy and atomic
current of a Bloch state, 
we show that such a hopping rate must be replaced by an effective hopping rate parameter
containing the additional contribution an interaction-driven hopping rate. 
Such a contribution turns out to be crucial at high filling factors, since it
preserves the positivity of the effective hopping rate parameter.
Level crossings 
between the energies
per particle of a Wannier-like state and the superfluid ground state 
are interpreted as a signature of 
the transition to configurations with macroscopically occupied states
at each lattice site.

\end{abstract}

\maketitle
\section{Introduction}
The study in the last decade of ultra-cold bosonic atoms in optical lattices has enabled 
the realization of 
an active and fruitful convergence
of atomic and condensed matter physics. Particularly, the analogy of such systems
with a solid material, where
the bosons play the role of the superconducting electron pairs and the laser beams act as
the ionic crystal, became the leitmotiv of numerous applications \cite{bloch,yuka}. 
In their seminal experiment, Greiner {\it et al.} \cite{greiner} 
showed that by increasing the lattice potential depth in a three-dimensional
optical lattice,
a quantum phase transition from a superfluid state to a Mott
insulating state can be achieved. This had been predicted by Jaksch {\it et al.} \cite{jak},
who accurately described such a transition within a Bose-Hubbard model at filling factors
of the order of unity. Actually, most research has so far been focused on optical lattices
with such a low filling factor, whereas the high filling factor domain appears scarcely 
treated. Such high-filling configurations are expected to be noticeably
affected by the on-site interaction between bosons, as 
the Wannier single-particle ground-state wave function in every site should be replaced by
a macroscopic wave function \cite{stoof}.
A suitable  configuration to experimentally investigate this type of condensates
 could be given by a ring-shaped lattice, where
a toroidal trap becomes symmetrically divided by a number of potential barriers radiating away from the
trap center \cite{pra10}. In fact, apart from presenting the ideal geometry to
sustain persistent currents, such a lattice would also exhibit a perfect azimuthal periodicity
for any number of lattice sites. This would permit to achieve extremely high
filling factors within the present experimental possibilities for the maximum number
of particles in the whole condensate. 
The effect of raising a single barrier across a long-lived persistent current in a toroidal condensate,
has recently been investigated as the first realization of an elementary closed-loop atom circuit \cite{rama}.
The generalization of such experiments to  ring lattices has shown
to be quite attainable in the light of the works of Amico {\it et al.} \cite{amico}
and Henderson {\it et al.} \cite{hend}. In fact, while
the former have thoroughly discussed
 the experimental setup for realizing a ring lattice,
such a system was actually
generated by the latter, utilizing a
rapidly moving laser beam that `paints' a time-averaged optical dipole potential,
transforming a toroidal condensate into a ring lattice.

From a theoretical viewpoint, recent investigations have analyzed
the effect of rotation on the ground state properties of bosonic atoms confined in a
one-dimensional ring lattice at low filling factors \cite{rey}.
A nonrotating ring lattice, on the other hand, has been predicted to
sustain persistent currents \cite{case,*dunn} if the phase 
difference between adjacent sites takes certain values \cite{pra10}.
In addition, the buildup of winding number in the
phase transition from Mott insulator to superfluid driven by tunneling rate increase,
has been shown to proceed through the so-called
Kibble-Zurek mechanism, except for very slow quench times \cite{zur1,*zur2}.
In the present work we will concentrate our attention 
on such nonrotating configurations with high barriers and high filling factors.
The starting point of a theoretical approach to
this kind of systems 
should consist in exploring an adequate variant of the Bose-Hubbard (BH) model, 
which should be expected
to exhibit occupation dependent parameters \cite{hazz,dutta}. As usual,
the main ingredient to derive such a
 BH Hamiltonian consists in finding a suitable set of orthogonal Wannier-like
functions, for which a number of variational schemes have been proposed
 \cite{li,wu,vig,dutta,faust}. Here, rather than resorting to such methods,
we shall obtain our set of Wannier-like functions simply as a `basis change' from
the orthogonal set of stationary `vortex' Bloch states \cite{fer05,gar07}.
Then, it will be shown that such functions possess the main properties
of the single-particle Wannier functions, except for their dependence on the filling
factor, 
and thus they become the adequate 
tool to study the slightly perturbed Bloch states arising from small occupation number imbalances,
or from small changes on the relative phase between
adjacent sites. Under such conditions, 
a generalized BH Hamiltonian that takes into account
interaction-driven tunneling processes will be derived. 
Such contributions, which were previously investigated for double- and triple-well configurations
\cite{anan,jia,viscondi}, will be shown to play a principal role at high filling factors.
Finally, by considering 
the level crossing between the energies per particle of a Wannier-like state and the superfluid ground
state, we will discuss the transition to configurations with a macroscopic occupation at each
site.  

This paper is organized as follows. In Sec. II, we describe the ring lattice and
remaining condensate parameters. In Sec. III, we analyze the main properties of
Bloch and Wannier-like states. In Sec. IV, we derive the generalized BH Hamiltonian
from which the energies of Bloch states are calculated, and the continuity equation
at a given lattice site and the corresponding atomic current are extracted.
Finally, in Section V we discuss our numerical results for the tunneling parameters and
level crossings, while in Sec. VI we present our summary and
main conclusions. 

\section{Ring-shaped lattice and condensate parameters}\label{descrip}
 
We consider  a Bose-Einstein condensate  of rubidium atoms confined by 
an external trap $ V_{\text{trap}}$, consisting of a superposition of
a toroidal term $V_{\text{toro}}$  and a lattice potential $V_{\text{L}}$
 formed by radial barriers.
Similarly to the trap 
utilized in  recent experiments \cite{ryu07,wei08}, the toroidal trapping  potential
 in cylindrical coordinates reads,
\begin{equation}
V_{\text{toro}}(r,z ) = \frac{ M }{2 } \left[\omega_{r}^2  r^2 
+ \omega_{z}^2  z^2\right] +  
V_0 \, \exp ( -2 \, r^2/\;\lambda_0^2)
\end{equation}
where $\omega_{r}$  and $\omega_{z}$ denote the radial and axial 
frequencies, respectively, and $M$ denotes the atom mass.
We have set $\omega_z >>
\omega_r$ to suppress excitation in the $z$ direction.
In particular, we have chosen
 $ \omega_r / (2 \pi) =  7.8 $ Hz  and $ \omega_z / (2
\pi) = 173 $ Hz, while for the laser beam we have set 
$ V_0  =  100  \, \hbar \omega_r$  and $ \lambda_0 = 6 \, l_r $, with 
$ l_r = \sqrt{\hbar /( M \omega_r)}$. 
On the other hand, the lattice  potential is formed by $N_c$
Gaussian  barriers of width $\lambda_b$ and amplitude $V_b$,
 located at equally spaced angular positions $ \theta_k = 2 \pi k /N_c $,
where $-[[(N_c-1)/2]]\leq k\leq [[N_c/2]]$ with $[[\cdot]]$ denoting the integer part,
\begin{equation}
V_{\text{L}}(x,y) =
V_b \,\, \sum_{k=-[[(N_c-1)/2]]}^{[[N_c/2]]} \Theta[ \sin(\theta_k) \, y  + \cos(\theta_k) \,x]
\,\,\,\,\exp \left\{ - \frac{ [ \cos(\theta_k) \, y  - \sin(\theta_k) \,x]^2}
{ \lambda_b^2}\right\},
\end{equation}
where $\Theta$ denotes the Heaviside function.

In the mean-field approximation, the stationary states are solutions of the Gross-Pitaevskii 
(GP) equation \cite{gross,*pita}
\begin{equation}
\left[-\frac{ \hbar^2 }{2 M}{\bf \nabla}^2  +
V_{\rm{trap}}({\bf r})+g\,N|\psi({\bf r})|^2\right]\psi({\bf r})
= \mu\,\psi({\bf r})
\label{gp}
\end{equation}
where $N$, $\mu$ and $ \psi({\bf r})$ respectively denote the number of particles,
the chemical potential and a two-dimensional (2D)
order parameter normalized to one \cite{castin}. 
The effective 2D coupling constant $g=g_{3D}\sqrt{M\omega_z/2\pi\hbar} $
  is written in terms of the 3D coupling constant between the atoms 
$g_{3D}=4\pi a\hbar^2/M$, where $a= 98.98\, a_0 $ denotes the  
$s$-wave scattering length of $^{87}$Rb, $a_0 $ being the Bohr radius.

\section{ Bloch and Wannier-like states}

We shall restrict our treatment to the case of 
high enough barrier heights, where quantum tunneling
between sites turns out to be the dominant dynamical process.
Such a regime arises when
the ground-state chemical potential becomes smaller than
the minimum of the effective potential barrier dividing two lattice sites
\cite{pra10}.
The most general solution of the GP equation (\ref{gp})  is given by
a Bloch state of the form \cite{fer05,gar07}
\begin{equation}
\psi_m( r, \theta ) = e^{i m \theta } \, f_m ( r, \theta ) \,,
\label{wfbloch}
\end{equation}
where  $ f_m ( r, \theta )$ is invariant under rotations in $ 2 \pi /  N_c $ and the winding number $m$ 
plays the role of an `angular' pseudomomentum
satisfying the
constraint
$-[[(N_c-1)/2]]\leq m\leq [[N_c/2]]$.
Such a constraint arises from the fact that all possible solutions can be reduced to those existing
in the first Brillouin zone in pseudomomentum space \cite{gar07}. We shall restrict ourselves
to Bloch states of the lowest energy, i.e., to the ground `vortex' states \cite{pra10}. 
In the language of crystal lattices, we would say
that we shall restrict our treatment to 
  the subspace of Bloch states of the `ground band'. In such a context,
the orthogonality of a pair of
 Bloch states,  $\psi_m$ and $\psi_n$,
can be easily proven as follows. First, the corresponding integral may be split into
separate integrals over each site, where we make the change of variable $\theta'=\theta-\theta_k$. 
Then, taking into account 
the rotational symmetry of the corresponding functions $f_m$ and $f_n$,
along with the equality
\begin{equation}
\sum_k \exp[i(m-n)\theta_k]=\delta_{m,n}N_c\,,
\label{suma}
\end{equation}
the orthogonality can be demonstrated.

Now, taking into account the 
periodicity of a Bloch state in the reciprocal lattice, $\psi_{m+jN_c}=\psi_m$,
it must have a Fourier series expansion with `wave vectors' $\theta'_k=(\theta_k+\theta_{k+1})/2=
\theta_k+\pi/N_c$ in  the direct lattice as follows \cite{ash}\footnote{Note in Eq. 
(\ref{fourier}) that the phase factor $\exp(-im\pi/N_c)$ 
has been absorbed into the expression of the Bloch wave function.},
\begin{equation}
\psi_m({ r, \theta })=\frac{1}{\sqrt{N_c}} \sum_{k} w_k({ r, \theta})
 \, e^{i \theta_k m}\,,
\label{fourier}
\end{equation}
where the Fourier coefficients in (\ref{fourier}) are given by the inversion formula
\footnote{The inversion formula (\ref{wannier}) can be readily checked by taking into account Eq.
(\ref{suma}).}
\begin{equation}
w_k({ r, \theta })=\frac{1}{\sqrt{N_c}} \sum_{n} \psi_n({ r, \theta})
 \, e^{-i n\theta_k } \,,
\label{wannier}
\end{equation}
with the summation over the angular pseudomomentum $n$ being restricted to the first Brillouin zone.
Replacing (\ref{wfbloch}) in (\ref{wannier}) and taking into account the symmetry of $f_n$,
we may realize that the Fourier coefficients arise from a single function $w(r,\theta)$ as follows,
\begin{equation}
w_k(r, \theta )= w( r, \theta-\theta_k)=\frac{1}{\sqrt{N_c}} \sum_{n} f_n( r, \theta-\theta_k)
 \, e^{i  n( \theta-\theta_k)} \,.
\label{wffn}
\end{equation}
Thus, pushing forward with the analogy to crystal lattices, we could name the function
\begin{equation}
w( r, \theta)=w_0( r, \theta)=\frac{1}{\sqrt{N_c}} \sum_{n} \psi_n( r, \theta),
\label{wan}
\end{equation}
the `Wannier' function of the ground band \cite{ash}.
Although we shall see that it shares many formal properties with 
the well-known Wannier functions, we shall also show that it presents a remarkable difference.
So, we feel more appropriate to speak in the following of a {\it Wannier-like} function. 
Let us first show the similarities.
Taking into account that the Bloch `vortex' states \cite{gar07} fulfill $\psi_n^*=\psi_{-n}$,
while the ground  and  highest states, $\psi_0$ and  $\psi_{ N_c/2}$ ($N_c$ even), respectively, are
 real,
it is easy to show that $w( r, \theta)$ given by
 (\ref{wan}) must be a real function. 
Also, from the orthonormality of Bloch wave functions 
and Eq. (\ref{suma}), one may readily check that the set of Wannier-like
functions centered on different
$k$-sites, $w (r, \theta-\theta_k)$, form indeed 
an orthonormal basis of the subspace of Bloch states of the ground band.
In addition, given that the Bloch `vortex' states fulfill $\psi_n(r,-\theta)=\psi_n^*(r,\theta)$,
it is easy to show that $w(r,\theta)$ turns out to be an even function of $\theta$ for odd $N_c$.
\begin{figure}
\includegraphics{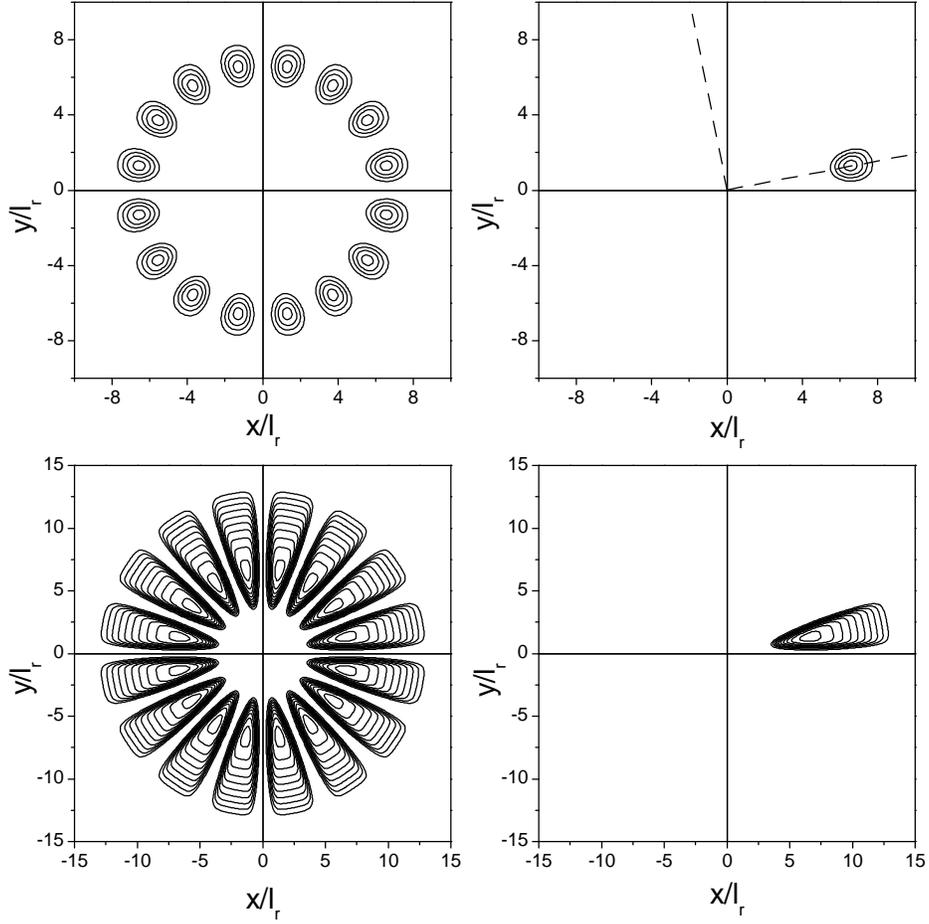}
\caption{Isocontours of the ground-state wave function density $|\psi_0|^2$ (left panels) and 
of the Wannier-like function density $w^2$ (right panels).
The  rotation of the coordinate system
 denoted by dashed lines in the top-right panel makes the Wannier-like function
symmetric with respect to the angular variable $\theta$.
The condensate parameters are $ V_b/\hbar \omega_r = 10$,
$N=10^3 $ (top), and
 $ V_b/\hbar \omega_r = 80$, $N=10^5 $ (bottom), while the number of lattice sites
and the Gaussian barrier width are given by $N_c=16 $ and $\lambda_b/ l_r =0.5 $, respectively.}
\label{wan1}
\end{figure}
On the other hand, by considering the rotation of the coordinate system in $\pi/N_c$ shown in Fig.~\ref{wan1},
which makes the Bloch wave function $\psi_{N_c/2}(r,\theta)$ an even function of $\theta$,
the same parity property may be readily extended to the case of $N_c$ even. Finally,
 by replacing the numerical solutions of the GP equation, $\psi_n({ r, \theta })$, 
in Eq. (\ref{wan}), we have shown that our Wannier-like
function is indeed a well-localized one, as seen in Fig.~\ref{wan1}.
However, there is a most remarkable difference between such a localized function and a
`true' Wannier function, which consists in that only the former turns out to depend on the
filling factor, i.e. the average number of particles at each site, as clearly observed in 
Fig.~\ref{wan1}. In fact, only for 
noninteracting bosons our Wannier-like function would not depend on the filling factor.
We have performed a calculation of 
the overlap between the Wannier-like function given by (\ref{wan})
  and the corresponding
 Wannier-like function for noninteracting bosons, i.e. with a vanishing coupling constant $g=0$,
which yielded 1.00, 0.98 and 0.48, for filling factors 5, 62.5
 and 6250, respectively. Particularly, the last two values correspond to the filling factors of
the top and bottom panels of Fig. 1, respectively. Therefore, we may conclude that only
for filling factors below $\sim 60$,
the Wannier-like functions should be almost independent of the average occupation number.
We will have more to say about this filling factor dependence in the following Sections.

To conclude it is instructive to rewrite Eq.~(\ref{fourier}) as
\begin{equation}
\psi_m({ r, \theta })=\frac{1}{\sqrt{N_c}} \sum_{k} w( r, \theta-\theta_k)
 \, e^{i m\theta_k }\,,
\label{localized}
\end{equation}
and notice that the above representation will be accurate to the extent that 
each site presents an almost uniform phase, which is consistent with 
a tight-binding scenario of high barriers
with a low particle current \cite{pra10}.

\section{ Bose-Hubbard model}\label{B-H} 

The above similarities between the Wannier-like function and the `true' Wannier functions, offer 
the adequate framework to 
establish a BH  model for our ring-shaped optical lattice.
As usual \cite{jak,blair,yuka},
the starting point is the second-quantized Hamiltonian
\begin{equation}
\hat H = \int d\,^2{\bf r}\,\, \hat{\Psi}^\dagger ({\bf r})\left[
-\frac{ \hbar^2 }{2 M}{\bf \nabla}^2  +
V_{\rm{trap}}({\bf r})\right]\hat{\Psi}({\bf r}) +
 \frac{g}{2}\int d\,^2{\bf r}\,\, \hat{\Psi}^\dagger ({\bf r})
\hat{\Psi}^\dagger ({\bf r})\hat{\Psi}({\bf r})\hat{\Psi}({\bf r}),
\label{hfield}
\end{equation}
where $\hat{\Psi}({\bf r})$ is the boson field operator.
We are interested in slightly perturbed Bloch states, which could be given by,
e.g., a small relative imbalance between the average population of two neighboring
sites. Then, for low enough temperatures, such configurations
will be conveniently described by expanding
the field operators in our Wannier-like basis of the ground band
\begin{equation}
\hat\Psi({\bf r}) = \sum_{k} w( r, \theta-\theta_k)\,\hat a_k,
\label{fieldw}
\end{equation}
where the operator $\hat a_k$ destroys a particle in the $k$-Wannier state
and satisfies the usual Bose commutation relations.
Here we remark that a possible dependence
of the operators $\hat a_k$  on the filling factor should, under the above conditions, be  negligible.
In fact, in the previous Section we have seen that this is actually the case for filling factors below 
$\sim$ 60, 
while for higher fillings, only
configurations that present small population imbalances
should be taken into consideration.

Then, replacing the field operators in (\ref{hfield}) through Eq. (\ref{fieldw}) and
assuming the tight-binding limit, where only the coupling to the nearest neighboring states
of any given Wannier-like state is taken into account, we obtain the following BH Hamiltonian
\begin{eqnarray}
\hat H_{BH} & = & \varepsilon\sum_k \,\hat a_k^\dagger\hat  a_k 
-J\sum_k (\hat a_k^\dagger\hat  a_{k+1}
+\hat a_{k+1}^\dagger\hat  a_k) \nonumber\\ & - &
\frac{J'}{2} \sum_k \left[ \hat a_k^\dagger\hat  a_k^\dagger\hat  a_k (\hat a_{k+1}+\hat a_{k-1})+
(\hat a_{k+1}^\dagger+\hat a_{k-1}^\dagger) \hat a_k^\dagger\hat  a_k \hat a_k\right]
 \nonumber\\ & + &
\frac{U}{2}\sum_k \hat a_k^\dagger\hat  a_k^\dagger\hat  a_k \hat a_k,
\label{hhb}
\end{eqnarray}
with
\begin{equation}
\varepsilon = \int d^2{\bf r}\,\, w( r, \theta)\left[
-\frac{ \hbar^2 }{2 M}{\bf \nabla}^2 +
V_{\rm{trap}}({\bf r})\right]w( r, \theta)
\label{eps0}
\end{equation}
\begin{equation}
J= -\int d^2{\bf r}\,\, w( r, \theta)\left[
-\frac{ \hbar^2 }{2 M}{\bf \nabla}^2  +
V_{\rm{trap}}({\bf r})\right]w( r, \theta\pm 2\pi/N_c)
\label{jota0}
\end{equation}
\begin{equation}
J'= -2\,g\int d^2{\bf r}\,\, w^3( r, \theta)
\,w( r, \theta\pm 2\pi/N_c)
\label{jotap0}
\end{equation}
\begin{equation}
U= g\int d^2{\bf r}\,\, w^4( r, \theta),
\label{U0}
\end{equation}
where the equivalence between the `$\pm$' expressions at the right-hand side of (\ref{jota0}) stems
from the reality of the Wannier-like functions, while 
the corresponding equivalence in (\ref{jotap0}) results from
the parity property of such functions.
In addition to the usual tunneling terms proportional to the standard
hopping rate $J$, we have also retained in (\ref{hhb}) interaction terms up to the first order in the
product of adjacent Wannier-like functions, which are proportional to the tunneling
parameter $J'$
\footnote{We have ignored in (\ref{hhb}) nearest-neighbor repulsion terms \cite{kuh},
because they are of
second order in the product of adjacent Wannier-like functions.
}. 
Later we will show that such interaction terms 
may constitute the most significant contribution to the tunneling rate at high filling factors.
The case of two sites $N_c=2$ is somewhat special since it 
is the only configuration presenting a single neighbor for each site.
Then, the expression (\ref{hhb}) reduces to
\begin{equation}
\hat H_{BH} = \varepsilon \hat N - \left[J+\frac{(\hat N-1)}{N_c}J'\right]
(\hat a_0^\dagger\hat  a_1+\hat a_1^\dagger\hat  a_0)
+\frac{U}{2}(\hat a_0^\dagger\hat  a_0^\dagger\hat  a_0 \hat a_0 +
\hat a_1^\dagger\hat  a_1^\dagger\hat  a_1 \hat a_1),
\end{equation}
where, for a fixed number of bosons $N$, 
the particle number operator $\hat N = \hat a_0^\dagger\hat  a_0+\hat a_1^\dagger\hat  a_1$
may be replaced by a $c$-number. Thus, we may see that 
the only difference with the standard two-mode BH Hamiltonian consists in that
the standard hopping rate $J$ is replaced by an effective hopping rate,
\begin{equation}
J_{\rm eff} = J+\frac{(N-1)}{N_c}J',
\label{jeff}
\end{equation}
which includes the additional contribution
of an interaction-driven hopping rate $\frac{N-1}{N_c}J'$ stemming from boson interactions.
Here it is worth noticing that an extended two-mode approach, which includes terms in the BH Hamiltonian
beyond the present approximation, has been recently investigated \cite{anan,gati}.

Next we obtain the mean value of the BH Hamiltonian
$\langle N,m|\hat H_{BH}|N,m\rangle$, where $|N,m\rangle$ represents
the quantum state of $N$ bosons condensed in the Bloch state (\ref{wfbloch}) of winding number $m$
\cite{fer05,gar07}.
To calculate such a matrix element, we may change to the Bloch basis
in (\ref{hhb}) by means of the expansion (cf. (\ref{wannier}))
\begin{equation}
\hat a_k^\dagger=\frac{1}{\sqrt{N_c}} \sum_{n} \hat \alpha_n^\dagger
 \, e^{-i n\theta_k } \,,
\end{equation}
where the operator $\hat \alpha_n^\dagger$ creates a particle in the corresponding Bloch state. 
Then, a straightforward calculation yields
\begin{equation}
E_m\equiv\langle N,m|\hat H_{BH}|N,m\rangle/N = \varepsilon + \frac{(N-1)}{N_c}\frac{U}{2}
-\nu J_{\rm eff}\cos(2\pi m/N_c),
\label{e_n}
\end{equation}
where $\nu$ denotes the number of neighbors ($\nu=2$ ($\nu=1$) for $N_c>2$ ($N_c=2$)).
We note that the above expression coincides with our previous result \cite{pra10} in the limit $N\gg 1$.

It is instructive to analyze the continuity equation for the $k$-th site of a lattice with $N_c>2$,
\begin{equation}
\frac{d}{dt}(\hat a_k^\dagger\hat a_k)=\frac{i}{\hbar}[\hat H_{BH},\hat a_k^\dagger \hat a_k]
=\hat J_{k-1\rightarrow k}-
\hat J_{k\rightarrow k+1},
\label{cont}
\end{equation}
where $\hat J_{k\rightarrow k+1}$ denotes the current operator for
atoms that move from site $k$ to site $k+1$,
\begin{equation}
\hat J_{k\rightarrow k+1}=\frac{i}{\hbar}(\hat a_{k+1}^\dagger \hat J_{\rm eff}^{(k)}\hat a_k-
\hat a_{k}^\dagger \hat J_{\rm eff}^{(k)}\hat a_{k+1}),
\label{corr}
\end{equation}
which has been written in terms of the hopping operator between sites $k$ and $k+1$
defined by
\begin{equation}
\hat J_{\rm eff}^{(k)}=J+\frac{J'}{2}(\hat a_k^\dagger \hat a_k+\hat a_{k+1}^\dagger \hat a_{k+1}).
\label{jefop}
\end{equation}
 The mean value of the current operator (\ref{corr})
for a condensate of $N$ particles in the Bloch state of winding number $m$ reads
\begin{equation}
\langle N,m|\hat J_{k\rightarrow k+1}|N,m\rangle=2 J_{\rm eff}\frac{N}{N_c}\sin(2\pi m/N_c),
\label{avcorr}
\end{equation}
which does not depend on the site we are considering, as expected. Note that analogously to
the mean value of the angular momentum \cite{pra10}, the current turns out
to be a sinusoidal function of the winding number. Note also its proportionality to the
effective hopping rate $J_{\rm eff}$, whereas for the standard BH model such a current turns out to be
 proportional to the
standard hopping rate $J$ \cite{pupi}.

The value of the BH model parameters (\ref{eps0})-(\ref{U0})
can be easily extracted from the mean-field energy of Bloch states ${\cal E}_m$ (see Appendix A).
Particularly, from the single value of energies of the ground state ${\cal E}_0$ and the highest excited state
${\cal E}_{N_c/2}$ ($N_c$ even) one obtains
\begin{equation}
\varepsilon = \frac{1}{2}({\cal E}_{N_c/2}^0+{\cal E}_0^0),
\label{eps}
\end{equation}
\begin{equation}
U = \frac{N_c}{N-1}({\cal E}_{N_c/2}^{int}+{\cal E}_0^{int}),
\label{U}
\end{equation}
\begin{equation}
J = \frac{1}{2\nu}({\cal E}_{N_c/2}^0-{\cal E}_0^0)
\label{jota}
\end{equation}
\begin{equation}
J' = \frac{N_c}{2\nu(N-1)}({\cal E}_{N_c/2}^{int}-{\cal E}_0^{int})
\label{jotap}
\end{equation}
where, according to Appendix A,
 the superscripts `$int$' and `0' denote interacting and noninteracting contributions to the energy,
respectively.

According to the Hamiltonian (\ref{hhb}), the energy per particle in a Wannier-like state,
i.e., neglecting tunneling processes, is given by
\begin{equation}
E_W=\varepsilon+\frac{U}{2}\left(\frac{N}{N_c}-1\right).
\label{Ew}
\end{equation}
It is interesting to compare
the above energy
to the energy per particle of the superfluid ground state 
$E_0$. Then, from Eqs.~(\ref{e_n}) and (\ref{Ew})
we obtain
\begin{equation}
E_W-E_0=\nu J_{\rm eff}-\left(\frac{N_c-1}{N_c}\right)\frac{U}{2}.
\label{dife}
\end{equation}
The existence of a superfluid to Mott insulator transition requires
the above difference to be positive for low barrier heights (superfluid regime),
and negative for high barrier heights (Mott insulator state).
Thus, the level crossing at an intermediate barrier height 
arising from (\ref{dife}), should be representing a transition to configurations where the system is 
well described by $N_c$ macroscopically occupied states.
We may utilize the above expression to obtain
the value at such a level crossing, $\eta_{cr}$,
of the dimensionless scaling parameter \cite{blair,yuka}
\begin{equation}
\eta=\frac{U}{\nu J_{\rm eff}},
\label{eta}
\end{equation}
relevant to the superfluid to Mott-insulator transition.
Thus, assuming $E_W=E_0$ in (\ref{dife}), we obtain
\begin{equation}
\eta_{cr}=2N_c/(N_c-1).
\label{eta_c}
\end{equation}
We may compare the above result with theoretical estimates focusing on critical values
of the parameter $U/J$. In fact, it has been pointed out that the superfluid to Mott-insulator
transition in a one-dimensional BH model can be described by the (1+1)D $O(2)$ model, which
gives \cite{bloch}
\begin{equation}
(U/J)_{cr}=2.2\,\bar{n}
\label{blochest}
\end{equation}
for filling factors $\bar{n}\gg 1$. The above proportionality to the filling factor
is also predicted from
the mean-field Gutzwiller ansatz, which yields
$(U/J)_{cr}=2\,(\sqrt{\bar{n}}+\sqrt{\bar{n}+1})^2\simeq 8\,\bar{n}$
for $\bar{n}\gg 1$ \cite{rey}. However, we must recall 
 that mean-field theories only provide a qualitative analysis in 1D systems. 
We must also remark that the result (\ref{blochest})
arises from a BH Hamiltonian that does not take into account the contribution of the
interaction-driven tunneling terms proportional to $J'$. 
So, such an estimate should only be reliable
for $J_{\rm eff}\simeq J$, i.e., for $J\gg \bar{n}J'$ ($\bar{n}\gg 1$).
However, we shall see in the following Section that such conditions are difficult to
reach within our condensate parameters.
Moreover, we shall show 
that the standard hopping rate $J$ becomes negative above certain filling factor, which means
that the parameter $U/J$ should actually increase with the average occupation number until
becoming divergent and meaningless above such a filling factor.  

\section{Numerical results}\label{numerical} 

In the BH model for linear lattices 
it is common to measure energies in units of the {\it recoil energy}  $E_R=\hbar^2k_B^2/2M$,
where the Bragg momentum $k_B$ corresponds to a lattice potential of the form $\sim \sin^2(k_B\, x)$.
To adapt this definition to the present case, we first note that a lattice potential
$\sim \sin^2(N_c\, \theta/2)$ would have the required angular periodicity of $2\pi/N_c$.
Then, recalling that without barriers the excitation energy per particle of 
a Bloch state of angular pseudomomentum
$m$ reads $K m^2$, where \cite{pra10} 
\begin{equation}
K = \frac{\pi\hbar^2}{M} \int \frac{ 1}{ r} [\psi_0( r)]^2  dr,
\label{k}
\end{equation}
we may realize that our `recoil energy' should be written
\begin{equation}
E_R = K (N_c/2)^2.
\label{recoil}
\end{equation}
We have performed numerical simulations for
three particle numbers, $N=$ 80, $10^3$ and $10^5$;
given that the corresponding recoil energies
turned out to be 0.743 $\hbar\omega_r$, 0.740 $\hbar\omega_r$ and 0.713 $\hbar\omega_r$,
 respectively, showing figures
that approximate the harmonic energy quantum $\hbar\omega_r$, we decided, for the sake of simplicity,
to keep such a value as our energy unit in all cases.

\begin{table}
\caption{Level crossings arising from Eq.~(\ref{dife}), see text for explanation.}
\begin{ruledtabular}
\begin{tabular}{ccccccccc} 
$N_c$ & $N$  & $N/N_c$ &
 $V_b/\hbar\omega_r$ & $V_{\rm min}/\mu_0$ & $\mu/\mu_0$ & $\eta_{cr}$ & $V_b/\hbar\omega_r$
\cite{bloch} \\
\hline
16 & 80 & 5 & 10.4 & 1.29 & 1.05 & 2.17 & 14.1 \\
 & 10$^3$ & 62.5 & 15.4 & 1.39 & 1.08  & 2.18 & 30  \\
 & 10$^5$ & 6250 & 95.4 & 1.72 & 1.25 & 2.23   \\
\hline
8 & 80 & 10 & 4.65 & 1.11 & 1.01 & 2.29 & 8.2  \\
 & 10$^3$ & 125 & 10.1 & 1.23 & 1.03 & 2.30 & 15.2  \\
 & 10$^5$ & 12500 & 81.4 & 1.53 & 1.11 & 2.32 &  \\
\hline
4 & 80 & 20 & 2.95 & 1.06 & 1.00 & 2.23 & 5.8 & \\
 & 10$^3$ & 250 & 8.93 & 1.20 & 1.01 & 2.68 &  \\
 & 10$^5$ & 25000 & 77.1 & 1.47 & 1.05 & 2.48 &  \\
\end{tabular}
\end{ruledtabular}
\end{table}

We have numerically evaluated the BH parameters through Eqs.~(\ref{eps})-(\ref{jotap}) for
the above particle numbers and  three numbers
of lattice sites, $N_c=$ 16, 8 and 4. In Table I, we display our numerical estimates for the level crossings
 for the different condensates and a Gaussian barrier width $\lambda_b/ l_r =0.5 $.
Apart from the dependence of the barrier height parameter $V_b$, it is interesting to compare 
the minimum of the effective potential barrier dividing two lattice sites
$V_{\rm min}$ \cite{pra10}, with the ground-state chemical potential $\mu$.
Recall that in Ref. \cite{pra10} we have identified the lower bound of the
quantum tunneling regime as  $V_{\rm min}/\mu\simeq 1$.
To scale out the dependence of the barrier height for different particle numbers, we have
represented $V_{\rm min}$ and $\mu$ in units of the chemical potential at zero barrier $\mu_0$
for each particle number. Compare also the numerical estimates for 
$\eta_{cr}$ shown in Table I to those given by the expression  (\ref{eta_c}), namely
$\eta_{cr}=$ 2.13, 2.29 and 2.67 for $N_c=$ 16, 8 and 4, respectively. Here it is worthwhile
pointing out that a similar agreement was found for wider Gaussian barriers 
($\lambda_b/ l_r =1 $). Finally, the last column of Table I shows the critical estimate for the barrier
height parameter $V_b$ arising from Eq.~(\ref{blochest}).
The absence of data for filling factors above 125 corresponds to the negative values obtained for the hopping
rate $J$. Note also that such critical barrier heights 
turn out to be always higher than those of the fourth column, as
expected.

\begin{figure}
\includegraphics{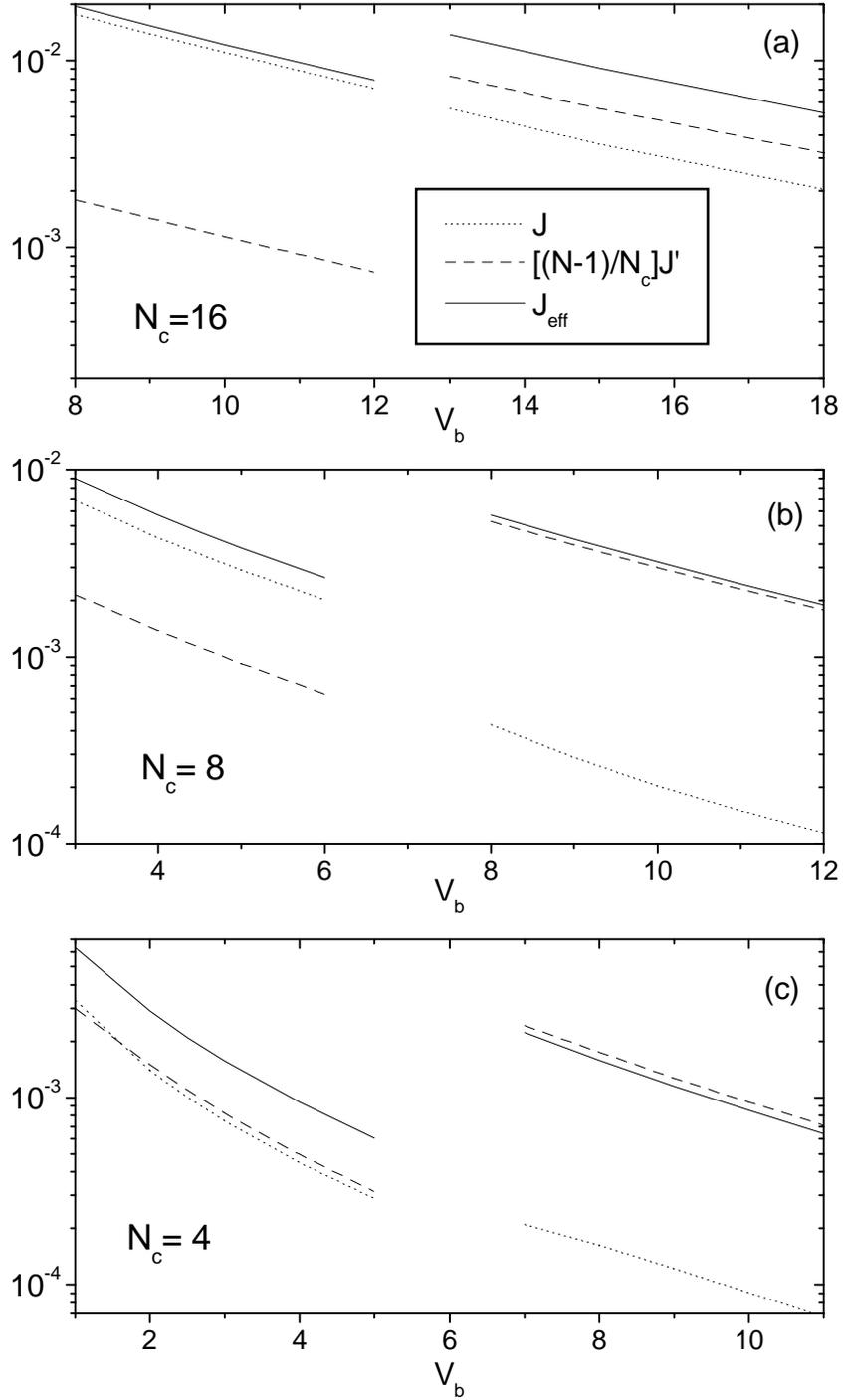}
\caption{Standard hopping rate $J$, 
interaction-driven hopping rate $\frac{N-1}{N_c}J'$, and effective hopping rate $J_{\rm eff}$,
as functions of the barrier height $V_b$ for the condensates of 80 particles (left)
and 10$^3$ particles (right). All quantities are given in units of $\hbar\omega_r$.
In panel (c) the standard hopping rate $J$ turns out to be 
negative for 10$^3$ particles (right), so we have
depicted its absolute value $|J|$. }
\label{fig2}
\end{figure}
\begin{figure}
\includegraphics{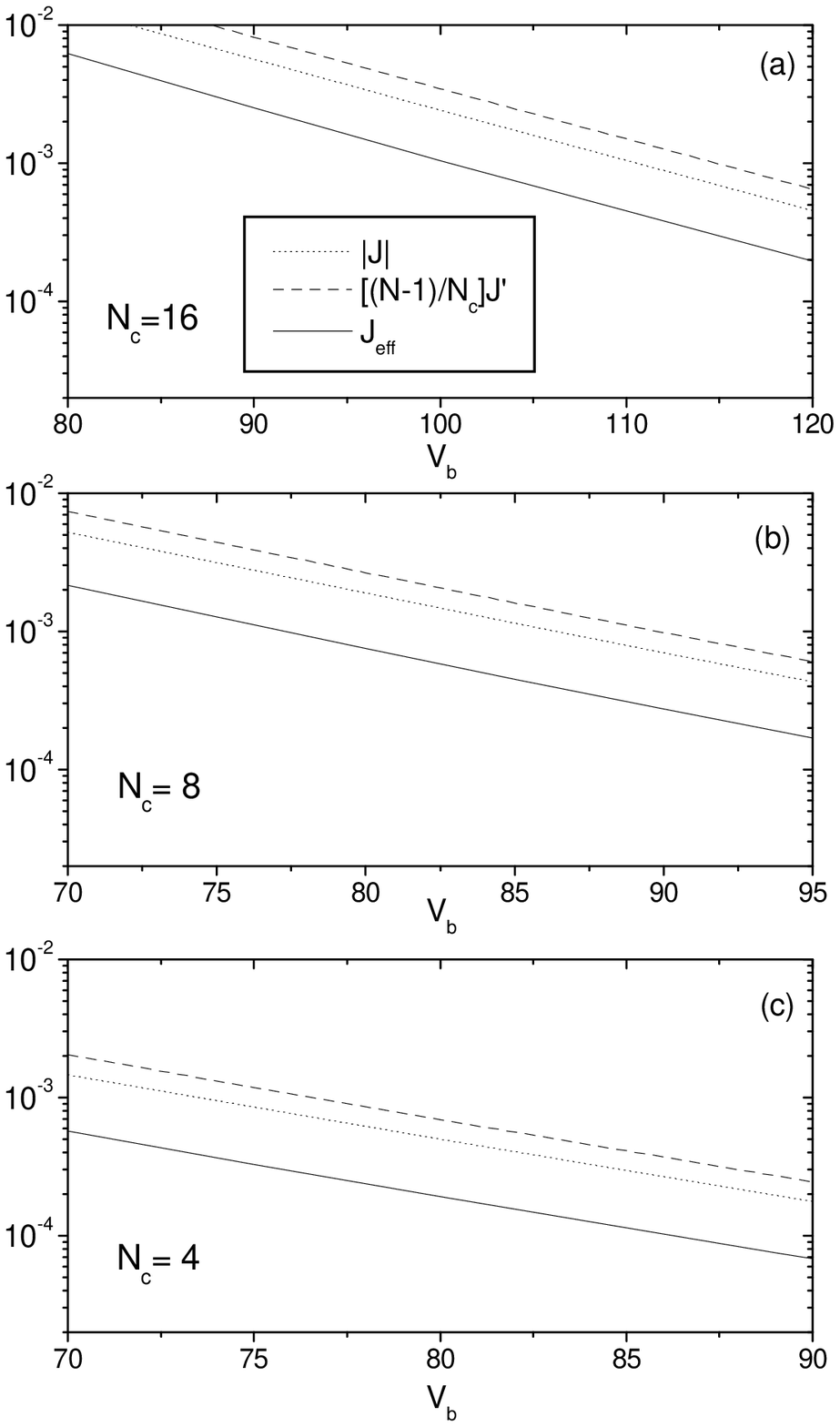}
\caption{Absolute value of the standard hopping rate $|J|$, 
interaction-driven hopping rate $\frac{N-1}{N_c}J'$, and
effective hopping rate $J_{\rm eff}$,
as functions of the barrier height $V_b$  for the condensate of 10$^5$ particles.
All quantities are given in units of $\hbar\omega_r$.
}
\label{fig8}
\end{figure}

In Figs.~\ref{fig2} and \ref{fig8} we depict the standard hopping rate $J$~(\ref{jota0})
computed from Eq.~(\ref{jota}), the 
tunneling parameter $\frac{N-1}{N_c}J'$~(\ref{jotap0})
computed from Eq.~(\ref{jotap}),
 and their sum $J_{\rm eff}$,
as functions of the barrier height.
Particularly, Fig.~\ref{fig2} shows that while the interaction component $\frac{N-1}{N_c}J'$
turns out to be almost negligible for 80 particles and $N_c=16$
(filling factor = 5),
for fewer lattice sites it shows a relative increase until becoming of the same order of the
hopping rate $J$ for $N_c=4$ (filling factor = 20). The larger filling factors of
$N=10^3$ yield an interaction component that turns out to be 
always larger than the standard hopping rate, to such an extent that
$J$ eventually becomes negative for $N_c=4$.
Note that such a dramatic change of sign occurs in between fillings
of 125 and 250 particles (Table I).
Negative values of $J$ were also predicted by Ananikian {\it et al.}
for large atom numbers in a double-well condensate \cite{anan}.
Finally, the extremely high fillings of $N=10^5$
yield again a negative $J$, as expected,
while a sort of saturation in the relative weights of the
interaction component and the negative hopping rate is observed.
This is reflected through the quite similar plots of Fig.~\ref{fig8},
despite the vertical shift for varying $N_c$, which 
arises from a decrease of the probability of
tunneling events as the number of barriers is lowered.

\begin{figure}
\includegraphics{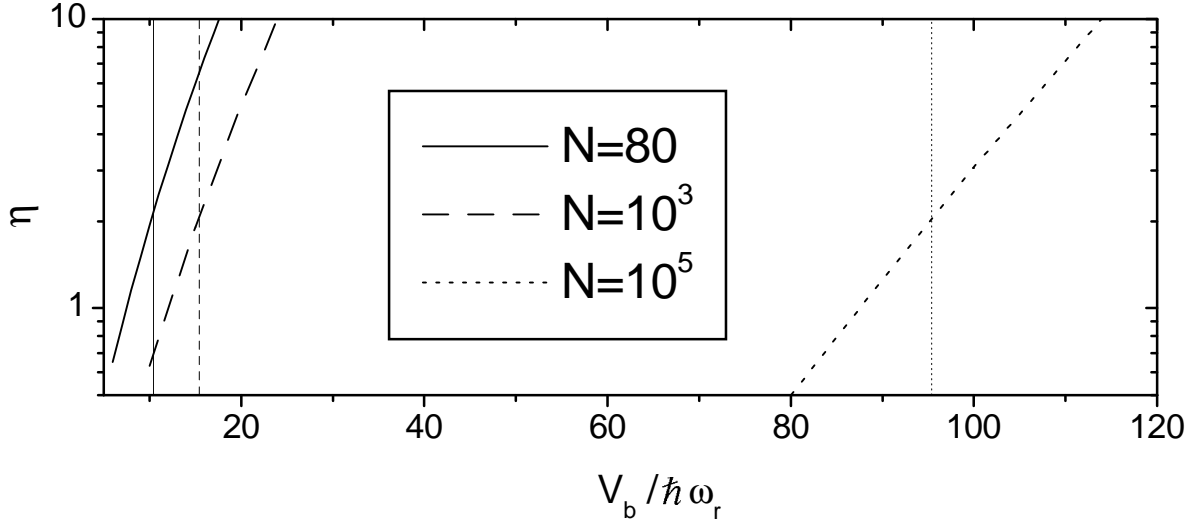}
\caption{Dimensionless scaling parameter $\eta=U/(\nu J_{\rm eff})$ as a function of
the barrier height $V_b$ for $N_c=16$ and the condensates of 80, 10$^3$ and 10$^5$ particles.
The vertical lines correspond to values in the fourth column of Table I.
}
\label{fig5}
\end{figure}
To conclude this Section, we display in Fig.~\ref{fig5}
the dimensionless scaling parameter $\eta=U/(\nu J_{\rm eff})$ versus the barrier height, for each 
number of particles and $N_c=16$. Notice that quite similar ranges of $\eta$ are obtained irrespective of
the barrier height interval, and this behavior repeats for the remaining values of $N_c$.

\section{ Summary and concluding remarks}\label{Conclusions} 
We have analyzed the high-barrier quantum tunneling regime of a Bose-Einstein condensate 
confined in a ring-shaped optical lattice. Representing
the orthogonal set of `vortex' Bloch states
through a basis of well-localized Wannier-like functions, we were able to formulate
a variant of the Bose-Hubbard model,
adequate for slightly perturbed Bloch states at any filling factor.
In addition to 
the usual hopping rate terms, such a Hamiltonian contains
interaction-driven tunneling terms,
which are shown to play its most important role when the standard hopping rate parameter becomes
negative at high filling factors. In fact, by calculating the energy and atomic
current of a Bloch state, 
we have shown that the standard hopping rate parameter must be replaced by an effective hopping rate
containing the additional contribution from the interaction-driven tunneling terms
in the BH Hamiltonian. 
We remark the importance of
such an interaction-driven hopping rate parameter, since it is
shown to preserve the positivity of the effective hopping rate
at high filling factors. A quite similar behavior for such hopping
rates was recently predicted for large atom numbers in a two-well configuration \cite{anan}.

We have found that, as the barrier height is increased,
the energies per particle of a Wannier-like state and the condensate ground state
exhibit a level crossing,  
which is interpreted as a signature of the transition to configurations with macroscopically occupied states
at each lattice site.
It is also shown that
the dimensionless scaling parameter, relevant to the superfluid to Mott insulator transition,
takes a remarkably simple expression
 at the level crossing, which only depends on the number of lattice sites.

Finally, we would like to point out that a future direction of the present studies will consist
in exploring the Boson Josephson-junction dynamics described by a generalized $N_c$-mode GP equation
\cite{anan,jia,[][{ in preparation.}]fut}.

\acknowledgments
DMJ and HMC acknowledge financial support from CONICET
under Grants Nos. PIP 11420090100243 and PIP 11420100100083, respectively.

\appendix

\section{Alternative calculation of the BH model parameters}\label{ap1}

An alternative calculation of the BH model parameters (\ref{eps0})-(\ref{U0}),
which avoids to change to a smaller numerical grid to deal with the tiny regions where the integrands
of the tunneling parameters (\ref{jota0}) and (\ref{jotap0}) are nonvanishing, proceeds as follows.
The method rests on the calculation of the mean-field energies of Bloch states ${\cal E}_m $,
which are obtained
by numerically solving the GP equation (\ref{gp}) for the order parameters $\psi_m$ \cite{pra10},
in order to evaluate the integral yielding the energy per particle
\begin{equation}
 {\cal E}_m   =\int \left( \frac{ \hbar^2 }{2 M}  |\nabla \psi_m |^2 +
V_{\rm{trap}} \,|\psi_m|^2 + \frac{1}{2} N g \, |\psi_m|^4  \right) dx\, dy,
\label{ed}
\end{equation}
where we may distinguish noninteracting and interacting contributions,
\begin{equation}
  {\cal E}^0_m   =\int \left( \frac{ \hbar^2 }{2 M}  |\nabla \psi_m |^2 +
V_{\rm{trap}} \,|\psi_m|^2  \right) dx\, dy ,
\label{ed0}
\end{equation}
and
\begin{equation}
  {\cal E}^{int}_m   =\int  \frac{1}{2} N g \, |\psi_m|^4   dx\, dy ,
\label{ed1}
\end{equation}
respectively.
In the context of this paper, i.e., for large barrier heights,
the above GP energies (\ref{ed0})-(\ref{ed1}) must coincide
 with those given by the corresponding contributions in (\ref{e_n})
with $J_{\rm eff}$ replaced from (\ref{jeff}).
Then, by calculating (\ref{ed0})-(\ref{ed1}) for
any two Bloch states and equating such results to the corresponding terms in (\ref{e_n}),
one can construct a linear system of four equations from which we may obtain the BH model parameters
(\ref{eps0})-(\ref{U0}). For instance,  the simplest choice for $N_c$ even corresponds to the 
winding numbers $m=0$ and $m=N_c/2$, which yields the following set of equations
\begin{equation}
{\cal E}_0^0 = \varepsilon-\nu J,
\label{a1}
\end{equation}
\begin{equation}
{\cal E}_0^{int} = \frac{(N-1)}{N_c}(U/2-\nu J'),
\label{a2}
\end{equation}
\begin{equation}
{\cal E}_{N_c/2}^0 = \varepsilon+\nu J,
\label{a3}
\end{equation}
\begin{equation}
{\cal E}_{N_c/2}^{int} = \frac{(N-1)}{N_c}(U/2+\nu J'),
\label{a4}
\end{equation}
 and the solution of such a system is given by the 
expressions (\ref{eps})-(\ref{jotap}).

Finally, a similar calculation for $N_c$ odd yields
\begin{equation}
\varepsilon = [{\cal E}_{(N_c-1)/2}^0+{\cal E}_0^0\cos(\pi/N_c)]/[1+\cos(\pi/N_c)],
\end{equation}
\begin{equation}
U = \frac{2N_c}{N-1}[{\cal E}_{(N_c-1)/2}^{int}+{\cal E}_0^{int}\cos(\pi/N_c)]/[1+\cos(\pi/N_c)],
\end{equation}
\begin{equation}
J = \frac{1}{2}[{\cal E}_{(N_c-1)/2}^0-{\cal E}_0^0]/[1+\cos(\pi/N_c)],
\end{equation}
\begin{equation}
J' = \frac{N_c}{2(N-1)}[{\cal E}_{(N_c-1)/2}^{int}-{\cal E}_0^{int}]/[1+\cos(\pi/N_c)].
\end{equation}

\providecommand{\noopsort}[1]{}\providecommand{\singleletter}[1]{#1}%

\end{document}